# The fourth generation, linac-ring type colliders, preons and so on

Saleh Sultansoy

TOBB University of Economics and Technology, Ankara, Turkey
ANAS Institute of Physics, Baku, Azerbaijan

## Abstract

Following a brief review of our contributions to the 2006 European PP Strategy, recent comments on several topics are presented. First of all, it is emphasized that only the simplest version of the fourth chiral generation, namely, minimal SM4 (mSM4) with only one Higgs doublet is in some tension with the recent LHC data on the Higgs boson search. This tension, which follows from the relative strengths of the H --> 4l and Higgs --> $\gamma\gamma$ channels, can be naturally resolved if there is a mechanism to enhance the H-->$\gamma\gamma$ width (although any charged and heavy particle could enhance the H-$\gamma$-$\gamma$ loop, 2HDM can be given as an example ).

We then emphasize the possible role of the linac-ring type colliders, especially LHeC (QCD Exploder) and TAC super charm factory. The QCD Explorer will give opportunity to enlighten the origin of the 98.5% portion of the visible universe's mass. Especially the $\gamma$-nucleus option seems to be very promising for QCD studies. The TAC super charm factory may provide opportunity to investigate the charm physics with statistics well above that of the dedicated runs at Super-B factories.

Finally, it is argued that the history of particle physics (and, more generally, the history of the investigation of the fundamental ingredients of the matter), a large number of "fundamental" particles, an inflation of observable free parameters, and, especially, the mixing of "fundamental" fermions favor the idea of a new set of fundamental particles at a deeper level. These new particles can be formulated with the preonic or even pre-preonic models.

**Contents:**
1. Introduction
2. The fourth generation: *"The rumors of my death have been greatly exaggerated, again!"*
3. Linac-ring type colliders:
    *i.* QCD Explorer to understand 98.5% of the visible universe mass
    *ii.* Cornerstone of the TAC project: Super-Charm factory
4. Preons or pre-preons
5. Conclusion: Importance of 'Energy Amplifier"
Appendix: Contents of Azerbaijani-Turkish contributions to the EPSS 2006:
    *i.* E. Arik and S. Sultansoy, "Turkish remarks …"
    *ii.* A.K. Ciftci and S. Sultansoy, "Linac-Ring Type Colliders"
    *iii.* S. Sultansoy, "SUSY, or not SUSY: that is the Question"
References

e-mail: ssultansoy@etu.edu.tr

# 1. Introduction

Establishment of the European Strategy Group was very important step for coordination and integration of HEP related research areas not only in Europe, but worldwide. Now we deal with the second stage of EPPS. Certainly, if properly organized, this stage will push forward High Energy Physics in foreseen future. At the same time it is very important to handle a mechanism for checking results of previous stage as well as of current stage in coming years. Our Azerbaijani-Turkish group made three contributions to EPSS 2006 (see Appendix for details).

I hope that success of the European Strategy Group (as well as ECFA) will positively influence other international bodies like ICFA etc. During the last decade I participated in two ICFA Seminars on "Future Perspectives in HEP" (in 2003 and 2011). Unfortunately, the decline was apparent: the first one was mach more democratic and creative. Science should stay far from extreme influence of different "scientific" lobbies.

## 2. The fourth generation: *"The rumors of my death have been greatly exaggerated, again!"*

After the recent discovery of the new particle with mass of 126 GeV by ATLAS and CMS, the fourth generation faced the third (may be most serious) trouble in its history. The first one, namely, the statement that according to the experimental results on invisible Z-decays the number of SM families should be 3, was rather curious. This was based on rudimentary belief: neutrinos are massless to provide V-A structure of weak interactions. The second one, namely, the statement that precision electroweak data excludes the fourth SM family, resulted from wrong interpretation of both precision electroweak data and SM4. Unfortunately, this wrong statement was included in PDG summaries and almost blocked SM4 activities for almost two decades. It should be noted that there are wide regions of SM4 points (analogs of the well-known SUGRA points), which are in better agreement with precision electroweak data than SM3 [1].

It should be mentioned that there are strong "democracy" arguments favoring the existence of the fourth SM family (see [2] and references therein). In addition, SM4 give opportunity to solve several tensions between SM3 and Nature. For these reasons a sequence of the SM4 workshops was organized since 2009 (see, for example, [3, 4]). As the result SM4 activities were essentially scaled up.

Concerning recent LHC results on the Higgs boson search, it should be emphasized that only the simplest version of the fourth chiral generation, namely, minimal SM4 (mSM4) with only one Higgs doublet is in some tension with the recent LHC data on the Higgs boson search. This tension, which follows from the relative strengths of the H --> 4l and Higgs --> $\gamma\gamma$ channels, can be naturally resolved if there is a mechanism to enhance the H-->$\gamma\gamma$ width (although any charged and heavy particle could enhance the H-$\gamma$-$\gamma$ loop, 2HDM can be given as an example). It is curious that both experiments, especially ATLAS, observe the deficit of di-photon events around 120 GeV. Therefore, situation in this channel may be different at the end of 2012.

In my opinion it is too early to start serious BSM activity on the modification of the Higgs boson properties (a lot of papers on the subject have been appeared in LANL arXive). I would suggest waiting till the end of 2012, when more data is expected to be available. Nevertheless, if there is a new physics at a scale of $\Lambda$, it should manifest itself in properties of heavy particles firstly. (*I never did believe in "Great Desert"*). Naively, one can expects that these manifestations are proportional to m/Λ or (m/Λ)$^2$. Obviously, contact interactions like $f_4$-$f_4$-V (V = g, W, Z, γ) can properly change H → VV branching ratios. At the same time new interactions can crucially change the search strategy for the fourth family fermions [2].

## 3. Linac-ring type colliders

Linac-ring type colliders have to mail goals: to explore TeV scale with lepton-hadron and photon-hadron collisions and to achieve highest luminosity at flavor factories (the history of corresponding proposals can be found in [5]. First goal is represented by the LHeC, which will explore the highest energy proton and ion beams available at the LHC (it seems that, concerning TeV scale exploration, LR type colliders have shot up from fourth to second place [6-8]). The second goal is represented by Super-Charm factory being developed within the framework of the Turkic Accelerator Complex project.

### *i.* QCD Explorer to understand 98.5% of the visible universe mass

QCD Explorer is the first (mandatory!) stage of the LHeC. The second (provisional) stage, namely, Energy Frontier can be realized only within the linac-ring option. Today, this option is considered as the basic one for the LHeC. Actually, keeping in mind reasons for the removal of the LEP from the tunnel, this decision was almost obvious from the beginning. Very important advantage of the linac-ring option is an opportunity to construct photon-hadron collider, too. It should be mentioned that this advantage disappear at the ERL based LHeC.

Details of the LHeC project can be found in [9]. Let me present here some personal comments. Using of one-pass linac could provide lumi of $10^{32}$cm$^{-2}$s$^{-1}$, which is quite enough for detailed QCD studies (for example, crucial investigation of small x region). Therefore, this can be considered as the first stage, followed by real gamma option as the second stage. Then, as the third stage, one can consider construction of a second one-pass linac for energy recovery (a la Litvinenko) to handle much higher lumi.

It should be emphasized that QCD Explorer will give opportunity to clarify the QCD basics at all levels from nucleus to partons and, therefore, to enlighten the source of 98.5% of the mass of the visible universe (comparing to 1.5% provided by the Higgs mechanism).

### *ii.* Cornerstone of the TAC project: Super-Charm factory

The title of this subsection is taken from the presentation at ECFA [10]. Although this statement met with objections from a few TAC related persons, it reflects the objective reality. At least, LR type charm factory with SR source was starting point of the TAC project (see twenty years old paper [11]). Moreover, accelerator science activities in Turkey had been started with linac-ring type collider topics. These activities provided an opportunity to sign Collaboration Agreement between DESY and Ankara University in summer 1997, during the visit of Professor Bjorn Wiik. It should be noted that this was the first collaboration agreement between DESY and any Turkish organization. Linac-ring type collider topics formed the basis for cooperation in the field of accelerator physics between Turkey and CERN.

Unfortunately, Super-Charm factory related activity in Turkey was almost blocked since 2006. Whereas, (on the contrary) this activity should be intensified keeping in mind recent competition with traditional (ring-ring type) proposals. I strongly hope that this failure is addressed as soon as possible.

**4. Preons or pre-preons**

Let me emphasize the analogy of today's SM fermions and parameters inflation with chemical elements inflation in 19th century and hadron inflation in 1950−1960. The both cases have been clarified through four stages: systematics, predictions confirmed, clarifying experiments, new basic physics level (see Table below). The last row to the Table is added in order to reflect present situation in particle physics.

Table: Historical analogy

| Inflation | Systematic | Confirmed Predictions | Clarifying experiments | Fundamentals |
|---|---|---|---|---|
| Chemical Elements | Mendeleyev Periodic Table | New elements | Rutherford | p, n, e |
| Hadrons | Eight-fold Way | New hadrons | SLAC DIS | Quarks |
| SM fermions | Flavor Democracy? | Fourth family ? | LHC ? | Preons ? |

Unfortunately, possible new structural levels of the matter (during my discussion with Professor Muhammed Abdus SALAM in autumn 1989, he emphasized that we need two levels at least) do not attract the necessary attention of HEP phenomenology scientists. I am sure that even 1% of efforts spent on SUSY-SUGRA related activities will result in essential progress in the preonic side.

**5. Conclusion: Importance of 'Energy Amplifier"**

Design of GeV energy, mA current proton accelerator for ADS applications should be among the priorities of our accelerator physicists. If HEP community contributes significantly to *green nuclear energy*, we can much more easily obtain the necessary resources to our field. An impact of this contribution to the mankind prosperity and sustainable development may exceed well the impact of WWW, PET and so on. Of course, it is quite possible that in foreseen future HEP will revolutionize the science, technology, economy and society in a manner similar to that happened one century ago.

**Appendix: Contents of Azerbaijani-Turkish contributions to the EPSS 2006.**

EPSS 2006 Breefing Book, Volume 2, Chapter 2 "Input from the community"
http://council-strategygroup.web.cern.ch/council-strategygroup/

*i.* **E. Arik and S. Sultansoy**, Turkish remarks on "Future Perspectives in HEP".
*1. Periodic Table of Elementary Particles; 2. Flavour Democracy → the Fourth SM Family; 3. SUSY vs Compositness → SUSY at pre(pre)onic level; 4. TeV scale lepton-hadron and photon-hadron colliders; 5.Linac-Ring type factories: TAC Project.*

*ii.* **A.K. Ciftci and S. Sultansoy**, Linac-Ring Type Colliders.

*1. Introduction; 2. ep Colliders: 2.1. QCD Explorer, 2.2. Energy Frontiers; 3. Additional eA, γp, γA and FEL γA options; 4. Particle Factories; 5. Suggestions.*

***iii.* S. Sultansoy**, SUSY, or not SUSY: that is the Question.
I. INTRODUCTION
II. PHENOMENOLOGY: II.1. Standard Model; II.2. Standard extensions: i) Higgs sector, ii) Fermion sector, iii) Gauge sector (up to GUT); II.2. Radical extensions i) Compositness (Preonic models): ii) SUSY (up to SUGRA), iii) Extra dimensions; II.3. Exotics
III. EXPERIMENT: III.1. Energy frontiers: i) Hadron colliders, ii) Lepton Colliders, iii) Lepton-hadron colliders; III.2 Particle factories; III.3. Fixed target, III.4. Non-accelerator
IV. CONCLUSİON